# Statistical Mechanical Foundations for Systems with Nonexponential Distributions

A. K. Rajagopal[1] and Sumiyoshi Abe[2]

[1]*Naval Research Laboratory*, *Washington D. C.*, *20375-5320*

[2]*College of Science and Technology*, *Nihon University*, *Funabashi*,

*Chiba 274-8501*, *Japan*

**Abstract**.     Traditionally the exponential canonical distributions of Gibbsian statistical mechanics are given theoretical justification in at least four different ways: steepest desecent method, counting method, Khinchin's method based on the central limit theorem, and maximum entropy principle of Jaynes. Equally ubiquitous power-law canonical distributions are shown to be given similar justification by appropriately adopting these formulations.

This is a report on our collaborative work done in the recent past [1-5] on an inquiry into the microcanonical foundation for power-law canonical distributions that occur so often in many physical contexts, in much the same way as the exponential canonical distributions of Gibbs do. There are at least two ways of understanding the exponential canonical distributions based solely on the principle of equal *a priori* probability: steepest descent and counting methods. The third way is due to Khinchin [6] based on the central limit theorem. Finally, the fourth is maximum entropy principle of Jaynes [7]. The third method is perhaps the most satisfactory mathematically, because only probabilistic principles are employed in establishing statistical mechanics of systems. In some way, the latter two methods have also the principle of equal *a priori* probability implied in their foundation.

We elaborate on the procedures needed to deduce the power-law canonical distributions. To examine the possibility of obtaining the power-law-type distribution, the form of the function we consider is

$$e_q(x) \equiv \left[1 + (1-q)x\right]^{1/(1-q)}, \tag{A}$$

which will be referred to as the *q*-exponential function. This is the inverse function of the *q*-logarithmic function

$$\ln_q(x) = \frac{x^{1-q} - 1}{1-q}. \tag{B}$$

Throughout this paper, $q$ is assumed to satisfy the condition

$$q > 1. \tag{C}$$

In the limit $q \to 1$, $e_q(x)$ and $\ln_q(x)$ converge to the ordinary exponential and logarithmic functions, respectively. For large $x$, the *q*-exponential function asymptotically exhibits the power-law behavior

$$e_q(x) \approx \frac{1}{x^{1/(q-1)}}. \tag{D}$$

Our aim here is to obtain this form for the canonical distribution function by various methods of statistical mechanics mentioned above.

**Steepest Descent Method**

Consider a classical system $s$ and take its $N$ replicas $s_1, s_2, \mathrm{L}, s_N$. The collection $\mathrm{S} = \{s_\alpha\}_{\alpha=1,2,\mathrm{L},N}$ is referred to as a supersystem. Let $A_\alpha$ be a physical quantity associated with the system $s_\alpha$. This is a statistical random variable and its value is denoted by $a(m_\alpha)$, where $m_\alpha$ labels the allowed configurations of $s_\alpha$. The quantity of interest is the average of $\{A_\alpha\}_{\alpha=1,2,\mathrm{L},N}$ over the supersystem: $(1/N)\sum_{\alpha=1}^{N} A_\alpha$. In microcanonical ensemble theory, the probabilities of finding $\mathrm{S}$ in the configurations in which the values of the average quantity lies around a certain value $\bar{a}$ [8]:

$$|M| < \varepsilon, \tag{1}$$

where

$$M \equiv \frac{1}{N}\sum_\alpha a(m_\alpha) - \bar{a}, \tag{2}$$

$$\varepsilon \sim O(N^{-1-\delta}) \qquad (\delta > 0). \tag{3}$$

The condition in eq. (3) will be revisited later. We restate eq. (1) as the step function, $\theta(x) = 1 \, (x > 0), \, 0 \, (x < 0)$. The microcanonical probability $\mathrm{P}(m_1, m_2, ... m_N)$ associated with eq. (1) is

$$P(m_1, m_2, \ldots m_N) \propto \theta(\varepsilon - |M|). \tag{4}$$

Let us select $s_1$ as the objective system. The probability of finding it in the configuration $m_1 = m$ is given by

$$p(m) = \sum_{m_2, \ldots m_N} P(m, m_2, \ldots m_N). \tag{5}$$

We can represent the step function by

$$\theta(x) = \int_{\beta - i\infty}^{\beta + i\infty} d\phi \, \frac{e_q(\phi x)}{2\pi i \phi}, \tag{6}$$

as long as $\beta$ satisfies

$$1 - (q-1)\beta x_{max} > 0. \tag{7}$$

Equation (6) reduces to the traditional representation when $q \to 1$. The condition in eq. (7) will be discussed later in the context of the steepest descent approximation.

Now, using eq. (6), we have in the leading order in $N$

$$\theta(\varepsilon - M) = \int_{\beta - i\infty}^{\beta + i\infty} d\phi \, \frac{e_q(\phi \varepsilon)}{2\pi i \phi} \prod_{\alpha=1}^{N} e_q\left(-\phi \frac{1}{N}[a(m_\alpha) - \bar{a}]\right). \tag{8}$$

In obtaining this equation, the $q$-exponential function is factorized as $e_q(\phi[\varepsilon - M]) \approx e_q(\phi \varepsilon) e_q(-\phi M)$ and then $e_q(-\phi M)$ is further factorized into the product over the supersystem. Since both $|M|$ and $\varepsilon$ are of $O(N^{-1-\delta})$ with $\delta > 0$, such a manipulation will be justified in the subsequent discussion of steepest descent method in the large-$N$ limit (as well as in the section on the generalized central limit theorem). Using $\theta(\varepsilon - |M|) = \theta(\varepsilon - M) - \theta(-\varepsilon - M)$ and performing a change of the integration variable as $\phi \to N\phi$, we have

$$\theta(\varepsilon - |M|) = \int_{\beta^*-i\infty}^{\beta^*+i\infty} d\phi \, \frac{\sinh_q(N\phi\varepsilon)}{\pi i \phi} \prod_{\alpha=1}^{N} e_q\left(-\phi[a(m_\alpha) - \bar{a}]\right), \tag{9}$$

where $\sinh_q(x) \equiv [e_q(x) - e_q(-x)]/2$ and

$$\beta^* = \frac{\beta}{N}. \tag{10}$$

Let us examine the condition in eq. (7). It is rewritten as

$$1 - (q-1)\beta^* N |\pm \varepsilon - M|_{\max} > 0. \tag{11}$$

This is a rectangular distribution function with a very narrow support. In fact, $|\pm \varepsilon - M|_{\max}$ is a quantity of $O(N^{-1-\delta})$ with $\delta > 0$. Therefore, $\beta^*$ can be an arbitrary positive constant in the large-$N$ limit.

In the leading order in $N$, the probability is

$$p_q(m) = \sum_{m_2, \ldots, m_N} P(m, m_2, \ldots, m_N)$$

$$= \frac{1}{W} \int_{\beta^*-i\infty}^{\beta^*+i\infty} d\phi \, \frac{\sinh_q(N\phi\varepsilon)}{\pi i \phi} \, e_q\left(-\phi[a(m) - \bar{a}]\right)$$

$$\times \sum_{m_2, \ldots, m_N} \prod_{\alpha=2}^{N} e_q\left(-\phi[a(m_\alpha) - \bar{a}]\right)$$

$$= \frac{1}{W} \int_{\beta^*-i\infty}^{\beta^*+i\infty} d\phi \, \frac{\sinh_q(N\phi\varepsilon)}{\pi i \phi} \, \frac{e_q\left(-\phi[a(m) - \bar{a}]\right)}{Z_q(\phi)} \exp[N \ln Z_q(\phi)], \tag{12}$$

where $W$ is the number of possible configurations allowed by the condition in eq. (1) and

$$Z_q(\phi) = \sum_m e_q\left(-\phi[a(m) - \bar{a}]\right). \tag{13}$$

Also, $W$ is given by

$$W = \int_{\beta^*-i\infty}^{\beta^*+i\infty} d\phi \, \frac{\sinh_q(N\phi\varepsilon)}{\pi i \phi} \exp[N \ln Z_q(\phi)], \tag{14}$$

because of the normalization of $p_q(m)$. Using the real part $\beta^*$ of $\phi$, the steepest descent condition yields, with the superscripts "$(S)$" to indicate steepest descent method,

$$\frac{\partial Z_q^{(S)}}{\partial \beta^*} = 0, \tag{15}$$

which gives rise to

$$p_q^{(S)}(m) = \frac{1}{Z_q^{(S)}(\beta^*)} e_q\left(-\beta^*[a(m) - \bar{a}]\right), \tag{16}$$

$$\bar{a} = \sum_m P_q(m) a(m), \tag{17}$$

simultaneously, where $P_q(m)$ is given by

$$P_q(m) = \frac{[p_q(m)]^q}{\sum_m [p_q(m)]^q}. \tag{18}$$

In the field of thermodynamics of chaotic systems [9], $P_q(m)$ in the above equation is referred to as the escort distribution. It is another probability distribution associated with the original distribution $p_q(m)$. These results follow from eq. (A) and $de_q(x)/dx = [e_q(x)]^q$. Clearly, all the discussions become the ordinary ones in Gibbsian canonical theory with the exponential distribution function in the limit $q \to 1+0$. The distribution function in eq. (16) is seen to behave asymptotically as

$$p_q(m) \sim \frac{1}{[a(m)]^{1/(q-1)}}, \tag{19}$$

which exhibits the desired power law.

The steepest descent condition leads to the fact that the arithmetic mean of $\{A(m_\alpha)\}_{\alpha=1,2,\text{L},N}$ coincides with the generalized expectation value with respect to the escort distribution as in eq. (17).

It is worth pointing out the appearance of the Legendre transform structure [10] in this development. Let us introduce

$$\Gamma_q^{(S)}(\beta^*) \equiv \ln_q Z_q^{(S)}(\beta^*) - \beta^* \, \bar{a}. \tag{20}$$

The maximum of $Z_q(\phi)$ in eq. (13) is found at the stationary point along the steepest descent path:

$$\left.\frac{\partial Z_q^{(S)}(\beta^*)}{\partial \beta^*}\right|_{\bar{a}} = 0 \quad \Rightarrow \quad \left.\frac{\partial \Gamma_q^{(S)}(\beta^*)}{\partial \beta^*}\right|_{\bar{a}} = -\bar{a}. \tag{21}$$

In addition, we also have the equivalent result

$$\frac{\partial Z_q^{(S)}(\beta^*)}{\partial \beta^*} = \left[Z_q^{(S)}(\beta^*)\right]^q \left\{\sum_m [a(m) - \bar{a}][p_q(m)]^q\right\} = 0, \tag{22}$$

provided that $\bar{a} = \langle A(m) \rangle_q \equiv \sum_m P_q(m)\, a(m)$.

Next, we examine counting method.

**Counting Method**

The number $W(m)$ of configurations characterized by eq.(1) where we now isolate the $m$th state, is written as

$$\left| \frac{1}{N}[a(m) - \bar{a}] + \frac{1}{N} \sum_{\alpha=2,\ldots,N} [a(m_\alpha) - \bar{a}] \right| < \varepsilon. \tag{23}$$

An important point is that the evaluation of $W(m)$ is context dependent, in general. For example, it is known [1, 11] that the counting rule in the fractal space is not unique. In the present context, we explore the possibility of using the $q$-logarithm of the density of configurations satisfying eq. (23). Thus, the number of configurations, $Y_N(\bar{a})$, obeying the condition

$$\left| \frac{1}{N} \sum_{\alpha=2,\ldots,N} [a(m_\alpha) - \bar{a}] \right| < \varepsilon \tag{24}$$

behaves according to

$$\frac{1}{N} \ln_q Y_N(\bar{a}) \to S_q^{(C)}(\bar{a}), \tag{25}$$

where $S_q^{(C)}(\bar{a})$ will be identified below with the entropy in accordance with the definition of temperature which will be denoted by $\beta$. Here, the superscripts "$(C)$" are used to indicate counting method.

When $N$ is large, the required number of configurations, $W(m)$, is now expressed in terms of $Y_N(\bar{a})$ in the form:

$$\ln_q W(m) = \ln_q \left( Y_N \left( \bar{a} - \frac{1}{N}[a(m) - \bar{a}] \right) \right)$$

$$\cong \ln_q(Y_N(\bar{a})) - [a(m) - \bar{a}] \frac{\partial S_q^{(C)}(\bar{a})}{\partial \bar{a}}. \tag{26}$$

Let us define

$$\beta' \equiv [Y_N(\bar{a})]^{q-1} \beta \equiv [Y_N(\bar{a})]^{q-1} \frac{\partial S_q^{(C)}(\bar{a})}{\partial \bar{a}}. \tag{27}$$

Then, equation (26), upon using the relation in eq.(A), may be expressed as

$$\frac{W(m)}{Y_N(\bar{a})} = e_q(-\beta'[a(m) - \bar{a}]). \tag{28}$$

Since the total number of configurations is given by $W = \sum_m W(m)$, we have, upon defining

$$Z_q^{(C)}(\beta) = W/Y_N(\bar{a}) = \sum_m e_q(-\beta'[a(m) - \bar{a}]), \tag{29}$$

$$p_q^{(C)}(m) = \frac{W(m)}{W} = \frac{1}{Z_q^{(C)}(\beta)} e_q(-\beta'[a(m) - \bar{a}]). \tag{30}$$

Now, we turn our attention to the development based on the generalized central limit theorem.

**Method Based on Generalized Central Limit Theorem**

In the preceeding two sections, we considered an arbitrary discrete physical quantiy $A_\alpha$. Here, we treat a continuous quantity, the energy of the system as a representative, in order to make use of the limit theorems in probability theory of continuous variables. Given a system composed of a large number of subsystems with energies $\{\varepsilon_i = E_i/B_N > 0\}_{i=1,2,\text{L},N}$, where $B_N$ is a positive $N$-dependent factor to be determined subsequently, let us consider its total energy $E = \varepsilon_1 + \varepsilon_2 + \text{L} + \varepsilon_N$. Recall the ordinary central limit theorem. If each of $E_i$'s obeys a common distribution $f(E_i)$ with ordinary finite second moment, $\langle E_i^2 \rangle_1 = \int_0^\infty dE_i \, E_i^2 \, f(E_i)$, then its $N$-fold convolution, $B_N(f*f*\text{L}*f)(B_N E)$, approaches the Gaussian distribution in the limit of large $N$, where $(f*g)(x) \equiv \int_0^x dx' \, f(x-x') g(x')$. This property was exploited by Khinchin [6] to establish Gibbsian canonical ensemble theory within the framework of probability theory.

Here, we note that the power-law distributions have no finite ordinary moments. However, the generalized central limit theorem states that the $N$-fold convolution of such a distribution still converges to a limit distribution. The associated limit distribution is the Lévy-stable distribution, denoted here by $F_\gamma(E_i)$. Since the energies are bounded from below, it suffices to consider the Lévy distribution in the half space, whose characteristic function is known to be of the form [12]

$$\chi_\gamma(t) = \int_0^\infty dE_i \, e^{iE_i t} F_\gamma(E_i) = \exp\left\{-a|t|^\gamma \exp\left[i \, \text{sgn}(t) \frac{\theta \pi}{2}\right]\right\}, \tag{31}$$

where $a$ is a positive constant, $\gamma \in (0, 1)$ is the Lévy index, $\theta$ a constant satisfying $|\theta| \le \gamma$, and $\text{sgn}(t) = t/|t|$ the sign function of $t$. If each of the $E_i$'s obeys $F_\gamma(E_i)$, then their $N$-fold convolution $B_N(F_\gamma * F_\gamma * \text{L} * F_\gamma)(B_N E)$ has the same characteristic function as that of the original $F_\gamma(E_i)$, that is, $\chi_\gamma^{(N)}(t) \equiv [\chi_\gamma(t/B_N)]^N = \chi_\gamma(t)$, provided that $B_N$ is chosen to be

$$B_N = N^{1/\gamma}. \tag{32}$$

$F_\gamma(E_i)$ is found to exhibit the following power-law behavior for large values of $E_i$:

$$F_\gamma(E_i) \sim E_i^{-1-\gamma}. \tag{33}$$

As an example of the distribution without finite ordinary moments, let us consider a power-law distribution

$$f(E_i) = \frac{1}{z(\beta)} \left[ \xi(\beta) + E_i \right]^{-s} \quad (1 < s < 2), \tag{34}$$

$$z(\beta) = \frac{\xi^{1-s}(\beta)}{s-1}. \tag{35}$$

In the above, $\xi(\beta)$ is a positive constant depending on a parameter $\beta$. The range of $s$ is determined by requiring that $f(E_i)$ is normalizable and all of its ordinary moments be divergent. The logarithm of the characteristic function of the $N$-fold convolution of this distribution is found to be [4]

$$\ln \chi^{(N)}(t) = -a |t|^{s-1} \exp\left[ i \operatorname{sgn}(t) \frac{\theta \pi}{2} \right] + M_N. \tag{36}$$

Here, $L(\alpha)$, $a$, and $\theta$ are given by

$$L(\alpha) = \int_0^\infty \frac{dy}{y^{1+\alpha}} (e^{-y} - 1) < 0 \quad (0 < \alpha < 1), \tag{37}$$

$$a = (s-1) \xi^{s-1}(\beta) |L(s-1)| > 0, \tag{38}$$

$$\theta = 1 - s, \tag{39}$$

respectively. $M_N$ is a quantity, which vanishes in the limit $N \to \infty$. Comparison of eqs. (31) and (36) leads to the identification of the Lévy index to be

$$\gamma = s - 1. \tag{40}$$

To fully characterize the limiting distribution $F_\gamma(E)$, it is necessary to relate the constant $a$ to a certain statistical physical quantity. In contrast to Khinchin's discussion, we have no finite ordinary moments of the distribution. This forces us to seek a different approach to determine the constant $a$ appearing in eq. (36).

In spite of the fact that the ordinary first moment, $\langle E_i \rangle_1$, is divergent, the generalized first moment, $\langle E_i \rangle_q \equiv \int_0^\infty dE_i \, E_i \, P_q(E_i)$, defined in terms of the escort distribution [9] associated with $f(E_i)$

$$P_q(E_i) \equiv \frac{[f(E_i)]^q}{\int_0^\infty dE_i \, [f(E_i)]^q} \tag{41}$$

is finite if $q$ is chosen to be

$$q = 1 + \frac{1}{s}. \tag{42}$$

This is the minimal requirement for the finiteness of $\langle E_i \rangle_q$. The denominator in eq. (41) and $\langle E_i \rangle_q$ are calculated to be

$$c_q \equiv \int_0^\infty dE_i \, [f(E_i)]^q = \frac{(s-1)^q}{s} \xi^{1-q}(\beta), \tag{43}$$

$$u_q \equiv \langle E_i \rangle_q = \frac{\xi(\beta)}{s-1}, \tag{44}$$

respectively. Thus, the constant $a$ in eq. (36) [and eq. (38)] is given in terms of the physical quantity $u_q$, i.e., the generalized internal energy. From eqs. (43) and (44), we have

$$c_q = \frac{s-1}{s} u_q^{1-q}. \tag{45}$$

To connect these quantities with those of thermodynamics, $\beta$ and $u_q$ are regarded as a conjugate pair in the Legendre transform structure [10]. For this purpose, an entropy functional, $S_q[f]$, is defined such that

$$\frac{\partial S_q^{(G)}}{\partial u_q} \equiv \beta, \tag{46}$$

$$-\frac{\partial \Gamma_q^{(G)}}{\partial \beta} \equiv u_q, \tag{47}$$

where

$$\Gamma_q^{(G)} \equiv S_q^{(G)} - \beta u_q. \tag{48}$$

The superscripts "$(G)$" are used to indicate the quantities derived within the scope of the generalized central limit theorem. The relation between $\xi$ and $\beta$ appearing in eq. (34) can be established, once the form of the entropy $S_q^{(G)}[f]$ is specified. At this stage, we examine the following form for the entropy:

$$S_q^{(G)}[f] = \frac{1}{1-q}\left(\sigma^{1-q} c_q - 1\right), \tag{49}$$

where a scale $\sigma$ is introduced to make $S_q^{(G)}[f]$ dimensionless. Using eq. (46), we find

$$\beta = \frac{s}{s-1} \frac{u_q^{-q}}{\sigma^{1-q}}. \tag{50}$$

From eqs. (48) and (50), we ascertain the validity of eq. (47).

It may be mentioned that in recent papers [3,4] similar studies on convergence properties in both the full and half spaces have been studied.

**Principle of Maximum Tsallis Entropy**

Actually the entropies appearing in eqs. (25) and (49) and the distributions in eqs. (16), (30), and (34) are the Tsallis nonextensive entropy [13] and its optimal distribution under the constraint on the generalized first moment with respect to the escort distribution in eqs. (18) (41), respectively.

To complete our program, here we make contact of the above three derivations with nonextensive statistical mechanics. This is done by maximizing the Tsallis entropy [13]

$$S_q[p] = \frac{1}{1-q}\left\{\sum_m [p(m)]^q - 1\right\} \tag{51}$$

subject to the constraints on the normalization of the probability $p(m)$ and on the normalized $q$-expectation value of a physical quantity, $a(m)$ [14]:

$$\langle A \rangle_q = \frac{\sum_m a(m)[p(m)]^q}{\sum_m [p(m)]^q} = \bar{a}. \tag{52}$$

Introducing the Lagrange multiplier $\beta$ to take into account the constraint in eq. (52), we obtain the following results:

$$p_q^{(M)}(m) = \frac{1}{Z_q^{(M)}(\beta)} e_q\left(-\tilde{\beta}[a(m) - \bar{a}]\right), \tag{53}$$

$$Z_q^{(M)}(\beta) = \sum_m e_q\left(-\tilde{\beta}[a(m) - \bar{a}]\right), \tag{54}$$

$$\tilde{\beta} \equiv \frac{\beta}{c_q}, \tag{55}$$

$$c_q \equiv \sum_m \left[p_q^{(M)}(m)\right]^q = \left[Z_q^{(M)}(\beta)\right]^{1-q}. \tag{56}$$

Here, the superscripts "$(M)$" is used to indicate the quantities obtained from the principle of maximum Tsallis entropy. Note that

$$\frac{\partial Z_q^{(M)}(\beta)}{\partial \beta} = 0, \tag{57}$$

in view of eq. (52). We now define $\Gamma_q^{(M)}(\beta)$ by means of the Legendre transform

$$\Gamma_q^{(M)}(\beta) \equiv \ln_q\left(Z_q^{(M)}(\beta)\right) - \beta \bar{a}, \tag{58}$$

so that

$$\frac{\partial \Gamma_q^{(M)}(\beta)}{\partial \beta} = -\bar{a}. \tag{59}$$

It is straightforward to show that

$$\frac{\partial S_q^{(M)}[p_q]}{\partial \bar{a}} = \beta. \tag{60}$$

Note that eqs. (59) and (60) define a canonical pair, which is of central importance in thermodynamics. It is useful to note that the above results go over to the standard ones in Gibbsin canonical ensemble theory in the limit $q \to 1$.

In conclusion, we have shown that the power-law canonical distribution can be derived within the fundamental principles of statistical mechanics based on four different methods: steepest descent method, counting method, the generalized central limit theorem, and the principle of maximum Tsallis entropy.

We thank Professors Paolo Grigolini and Constantino Tsallis for kindly inviting us to the workshop. One of us (A. K. R.) acknowledges the partial support from the US Office of Naval Research. The other (S. A.) was supported by the GAKUJUTSU-SHO Program of College of Science and Technology, Nihon University. He thanks the warm hospitality of the Naval Research Laboratory, Washington, DC, which made this collaboration possible.